\documentclass[preprint,pra,superscriptaddress]{revtex4-1}

\usepackage[dvips]{graphicx} 
\usepackage{amsfonts}
\usepackage{amssymb}
\usepackage{amscd}
\usepackage{amsmath}    
\usepackage{enumerate}
\usepackage{epsfig}
\usepackage{subfigure}
\usepackage{xcolor}

\newcommand{\bra}[1]{\mbox{$\left\langle #1 \right|$}}
\newcommand{\ket}[1]{\mbox{$\left| #1 \right\rangle$}}

\begin{document}

\title{Alternative schemes for measurement-device-independent quantum key distribution}

\author{Xiongfeng Ma}
\email{xma@tsinghua.edu.cn}
\affiliation{Center for Quantum Information, Institute for Interdisciplinary Information Sciences, Tsinghua University, Beijing, China}
\affiliation{Center for Quantum Information and Quantum Control,\\
Department of Physics, University of Toronto, Toronto,  Ontario, Canada}
\affiliation{School of Electronic and Electrical Engineering, University of Leeds, Leeds, United Kingdom}

\author{Mohsen Razavi}
\email{m.razavi@leeds.ac.uk}
\affiliation{School of Electronic and Electrical Engineering, University of Leeds, Leeds, United Kingdom}

%

\begin{abstract}
Practical schemes for measurement-device-independent quantum key distribution using phase and path or time encoding are presented. In addition to immunity to existing loopholes in detection systems, our setup employs simple encoding and decoding modules without relying on polarization maintenance or optical switches. Moreover, by employing a modified sifting technique to handle the dead-time limitations in single-photon detectors, our scheme can be run with only two single-photon detectors. With a phase-postselection technique, a decoy-state variant of our scheme is also proposed, whose key generation rate scales linearly with the channel transmittance.
\end{abstract}

\maketitle

\section{Introduction}
Quantum key distribution (QKD) enables two remote parties to securely exchange cryptographic keys \cite{BENNETT:BB84:1984,Ekert:QKD:1991}. Despite the theoretically provable security of QKD protocols \cite{Mayers_01,LoChauQKD_99,ShorPreskill_00}, achieving security  with realistic devices is still a challenge \cite{KoashiPreskill_03,ILM_07,GLLP_04,TT_Thres_08,BML_Squash_08}. In fact, before any security proofs can be applied to practical scenarios, various device imperfections should be carefully examined. For example, the detector efficiency mismatch can be exploited by eavesdroppers to implement the efficiency mismatch attack \cite{MAS_Eff_06} or the time-shift attack \cite{Qi:TimeShift:2007,Zhao:TimeshiftExp:2008}. Lately, other imperfections, such as the detector's after-gate pulses and the dead time, have also been exploited in hacking strategies \cite{Lydersen:Hacking:2010,Wiechers:AftergateAttack:2011,Weier:DeadtimeAttack:2011,Jain:AttackExp:2011}. Although, in each case, certain counter-measures have been proposed \cite{Yuan:BlindAttack:2010,Yuan:AntiBrightAttack:2011}, to fully remove such attacks one must deal with their fundamental root, i.e., the detection efficiency loophole. In this paper, we build on recent progress on measurement-device-independent QKD (MDI-QKD) \cite{Lo:MIQKD:2012, Tamaki:MIQKD:2012} to propose alternative practical schemes resilient to detection loopholes, thence shielding out all the aforementioned attacks in QKD systems.

The security loopholes in QKD systems essentially stem from the existing issues in Bell's inequality tests. There are three major loopholes, corresponding to the three assumptions in Bell's inequality tests,
\begin{enumerate}
\item
locality loophole \cite{Bohm.Local.57}, which is related to the assumption that two test parties are spacelike-separated;

\item
efficiency loophole \cite{Pearle_Bell_70}, which is related to the fair-sampling assumption \cite{EffLoop_08}; and

\item
randomness (free-will) loophole, which is related to the assumption that measurement bases are chosen randomly.
\end{enumerate}

In the context of QKD, some of these loopholes have proved to be more harmful than the others. For instance, it is reasonable to assume that the information in the two legitimate parties of QKD, Alice and Bob, is protected from the eavesdropper, Eve. Thus, the locality loophole does not necessarily lead to hacking strategies. With recent developments in quantum random number generators \cite{Jennewein:QRNG:2000,Xu:QRNG:2012}, the randomness loophole may not introduce security issues either. The efficiency loophole, however, opens up to many quantum attacks. In fact, the aforementioned attacks all fall into this category.

One approach to overcome device imperfections is by using device-independent QKD (DIQKD) schemes \cite{MayersYao_98,AGM_Bell_06,Acin:DeviceIn:07,McKague:Selftesting:2010}. The underlying assumptions of security in these schemes are relaxed to only a few, such as no, or little, direct leakage of key information out of QKD users. Unfortunately, DIQKD schemes impose severe constraints on the required specifications for physical devices in use. For example, the tolerable error rate is $7.1\%$ and the minimum required transmittance is $92.4\%$ \cite{Pironio:DeviceIn:09}, which make its experimental demonstration extremely challenging.

In order to relax the above constraints, several detection-device-independent QKD schemes have been proposed \cite{MXF:DDIQKD:2012,Branciard:OneDIQKD:2012}. The main additional assumption is that the source is trustful. In practice, many QKD schemes use simple source setups, which can be monitored in real time \cite{ZQL_untru_08,Peng:Source:08}. The detection system, on the other hand, is more vulnerable to attacks \cite{MAS_Eff_06,Qi:TimeShift:2007,Zhao:TimeshiftExp:2008,Lydersen:Hacking:2010,Wiechers:AftergateAttack:2011,Weier:DeadtimeAttack:2011,Jain:AttackExp:2011}. In \cite{MXF:DDIQKD:2012}, for instance, a higher error rate of $11\%$ and a lower transmittance of $65.9\%$ are allowed. We emphasize that the scheme presented in \cite{MXF:DDIQKD:2012} uses a partial self-testing technique to overcome loopholes in detection. However, as pointed out in \cite{EffLoop_08}, the time-shift attack puts an ultimate bound of 50\% on the transmittance. This is due to the random bit assignment to no-click events in \cite{MXF:DDIQKD:2012}. Recently, Lo, Curty, and Qi proposed an MDI-QKD scheme \footnote{The two terminologies, detection-device independent and measurement-device independent, have the same meaning. In order to avoid possible confusion with the full device-independent case, we often use the term ``measurement-device independent" instead of ``detection-device-independent".} that is able to essentially avoid random bit assignments, hence going beyond the 50\% efficiency limit \cite{Lo:MIQKD:2012}. By relying on entanglement swapping techniques \cite{Rainer:EntSwap:2009} and reverse EPR schemes \cite{Biham:ReverseEPR:1996}, the MDI-QKD scheme \cite{Lo:MIQKD:2012} (see also \cite{Braunstein:MIQKD:2012}) can achieve similar performance to traditional QKD systems, while shielding out detection loopholes.

Thus far, three schemes for MDI-QKD have been proposed, two of which rely on phase encoding \cite{Tamaki:MIQKD:2012}, and the original one uses polarization encoding \cite{Lo:MIQKD:2012}. The latter requires polarization maintenance over the quantum channel, which makes its implementation over optical fibers challenging. The phase-encoding scheme I in \cite{Tamaki:MIQKD:2012} essentially follows the coherent-state QKD scheme without phase randomization \cite{Lo:Decoy:2005}, and its key rate decays quadratically with the channel transmission efficiency \footnote{In the security proof of the decoy-state scheme, the phases of quantum signals are assumed to be randomized \cite{Lo:Decoy:2005}. Scheme I in \cite{Tamaki:MIQKD:2012}, however, relies on non-random and identical phase reference for Alice and Bob \cite{ LoPreskill:NonRan:2007}. The application of the decoy-state scheme to this setting is still an open question.}. Scheme II in \cite{Tamaki:MIQKD:2012} relies on the relative phase between two weak pulses. In order to perform entanglement swapping, in \cite{Tamaki:MIQKD:2012}, this phase information is converted to polarization states before being measured by a set of four single-photon detectors. Both phase-encoding schemes require fast optical switches in the measurement unit. We remark that scheme I in \cite{Tamaki:MIQKD:2012} is more robust against some imperfections of the state preparation, which may offer benefits in certain practical situations. Note that proof-of-principle field tests of the MDI-QKD scheme were presented recently \cite{Rubenok:MIQKDexp:2012,daSilva:MIQKD:2012,Liu:MIQKDexp:2012}.

In this paper, we propose alternative phase-encoding schemes. Comparing to the one proposed in Ref.~\cite{Tamaki:MIQKD:2012}, our schemes do not require optical switches in the setups. Moreover, the proposed schemes can be implemented with only two single-photon detectors, which makes them even more cost effective. By introducing a proper postselection technique, in the two-detector setup, we can minimize the effects of the dead time. Our schemes do not require polarization maintenance, and, if implemented with single-photon sources, they do not require a phase reference between Alice and Bob. The decoy-state versions of our setup are, however, sensitive to the choice of encoding bases, and in some cases require a mutual phase reference. By introducing a phase-postselection technique, we can, however, reduce the error rate in the decoy-state protocols. Note that the security of the proposed phase-postselection technique needs to be further investigated.

The rest of this paper is organized as follows. In Sec.~\ref{Sec:Path:MIQKD}, we propose a path-phase-encoding MDI-QKD scheme with single-photon states. We simplify our setup in Sec.~\ref{Sec:Path:Time} and generalize it to coherent-state sources in Sec.~\ref{Sec:Path:Coherent}. We conclude the paper in Sec.~\ref{Sec:Path:Conclusion}.

\section{Single-photon MDI-QKD} \label{Sec:Path:MIQKD}
\begin{figure}[hbt]
\centering
\resizebox{8cm}{!}{\includegraphics{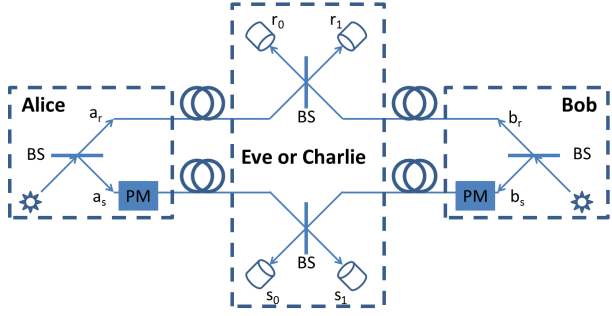}}
\caption{(Color online) A schematic diagram for the path-phase-encoding MDI-QKD scheme. Here, BS stands for 50:50 beam splitter and PM stands for phase modulator. Alice and Bob each encodes their qubits by introducing a relative phase shift between their reference and signal beams. The phase shifts are applied to the signal modes using PMs, chosen from the set $\{ 0 , \pi/2, \pi, 3 \pi /2 \}$. A partial BSM, possibly performed by an untrusted party, Eve or Charlie, on the two reference and the two signal modes would establish correlations between the raw key bits of Alice and Bob. Provided that they use the same phase basis, a joint click on detectors $r_0$ and $s_0$ implies identical bits for Alice and Bob, so does a joint click on $r_1$ and $s_1$. A joint click on $r_0$ and $s_1$, or, $r_1$ and $s_0$ would imply complement bits.}
\label{Fig:Path:MIdiag}
\end{figure}

In this section, we present an alternative MDI-QKD scheme using path and phase-encoding techniques \cite{Bennett:B92:1992,DLCZ_01, Sangouard:single-photon:2007}. A key component of our scheme is still a partial Bell-state measurement (BSM) module implemented by 50:50 beam splitters and single-photon detectors; see the Eve or Charlie's box in Fig.~\ref{Fig:Path:MIdiag}. In this section, we assume that perfect single-photon (qubit) sources are used by Alice and Bob. We use the setup in Fig.~\ref{Fig:Path:MIdiag} to illustrate how our scheme works. In Sec. \ref{Sec:Path:Time}, we will present a more practical setup for implementation purposes.

Our path-encoding MDI-QKD scheme works as follows. Alice and Bob, in Fig.~\ref{Fig:Path:MIdiag}, each prepares a single-photon state and passes them through 50:50 beam splitters. The resulting two modes are referred to as reference and signal modes, denoted, respectively, by $a_r$ and $a_s$ on Alice's side, and $b_r$ and $b_s$ on Bob's side. In order to generate the four states of the BB84 protocol, phase modulators, respectively, introduce relative phase shifts $\theta_a$ and $\theta_b$ between the reference and signal modes of Alice and Bob to obtain the following state
\begin{equation} \label{Path:MIQKD:OriAB}
\begin{aligned}
(\ket{1}_{a_r}\ket{0}_{a_s}&+e^{i\theta_a}\ket{0}_{a_r}\ket{1}_{a_s}) \otimes
(\ket{1}_{b_r}\ket{0}_{b_s}&+e^{i\theta_b}\ket{0}_{b_r}\ket{1}_{b_s}), \\
\end{aligned}
\end{equation}
where normalization factors are neglected for now. To follow the BB84 protocol, Alice and Bob randomly choose $\theta_a$ and $\theta_b$ from the two basis sets of $\{0, \pi\}$ and $\{\pi/2, 3\pi/2\}$. Phase values $0$ and $\pi/2$ represent bit 1 and the other two represent bit 0. When single-photon sources are used, the overall phase has no effect on the final result and will be neglected here.

To better understand how the setup in Fig.~\ref{Fig:Path:MIdiag} works, let us first neglect the channel loss and dark count effects, which will be addressed in Appendix \ref{App:Path:SinglePhoton}. We also assume that the relative phase between the reference and signal modes is preserved; we will see that, in the next section, how this can practically be achieved. A successful partial BSM in Fig.~\ref{Fig:Path:MIdiag} occurs when one, and only one, of $r_0$ and $r_1$, and one, and only one, of $s_0$ and $s_1$ click. All other detection events, such as the case when both $r_0$ and $r_1$ click, are discarded. Conditioned on a successful BSM outcome, the relevant terms in the joint state of Alice and Bob are given by
\begin{equation} \label{Path:MIQKD:ConAB}
\begin{aligned}
\ket{1}_{a_r}\ket{0}_{a_s}\ket{0}_{b_r}\ket{1}_{b_s} +e^{i(\theta_a-\theta_b)}\ket{0}_{a_r}\ket{1}_{a_s}\ket{1}_{b_r}\ket{0}_{b_s}. \\
\end{aligned}
\end{equation}
The above state will go through two 50:50 beam splitters, which remove any which-way information, in the BSM module resulting in
\begin{equation} \label{Path:MIQKD:ConABPhase}
\begin{aligned}
&\ket{01+10}_{r_0r_1}\ket{01-10}_{s_0s_1} +e^{i(\theta_a-\theta_b)}\ket{01-10}_{r_0r_1}\ket{01+10}_{s_0s_1} \\
&= \ket{0101-0110+1001-1010}_{r_0r_1s_0s_1} + e^{i(\theta_a-\theta_b)}\ket{0101+0110-1001-1010}_{r_0r_1s_0s_1}, \\
\end{aligned}
\end{equation}
where $r_0$, $r_1$, $s_0$ and $s_1$ represent the input modes to the corresponding detectors in Fig.~\ref{Fig:Path:MIdiag}, and we have used the following transformation for the two beam splitters:
\begin{equation} \label{Path:MIQKD:DetTrans}
\begin{aligned}
\ket{1}_{a_r}\ket{0}_{b_r} &\mapsto \ket{0}_{r_0}\ket{1}_{r_1} + \ket{1}_{r_0}\ket{0}_{r_1}, \\
\ket{0}_{a_r}\ket{1}_{b_r} &\mapsto \ket{0}_{r_0}\ket{1}_{r_1} - \ket{1}_{r_0}\ket{0}_{r_1}, \\
\ket{1}_{a_s}\ket{0}_{b_s} &\mapsto \ket{0}_{s_0}\ket{1}_{s_1} + \ket{1}_{s_0}\ket{0}_{s_1}, \\
\ket{0}_{a_s}\ket{1}_{b_s} &\mapsto \ket{0}_{s_0}\ket{1}_{s_1} - \ket{1}_{s_0}\ket{0}_{s_1}. \\
\end{aligned}
\end{equation}
In the above equation, we assume that the photons arriving at the relay are indistinguishable. This can be guaranteed by applying filters \footnote{The three main dimensions that might need filtering are polarization, frequency, and time.} before the 50:50 beam splitters in the measurement box. 

If $\theta_a-\theta_b=0$, then the state in Eq.~\eqref{Path:MIQKD:ConABPhase} becomes
\begin{equation} \label{Path:MIQKD:ConABdetc}
\begin{aligned}
\ket{0101-1010}_{r_0r_1s_0s_1};
\end{aligned}
\end{equation}
that is, either detectors $r_0$ and $s_0$, and only these two, click or $r_1$ and $s_1$ click. Otherwise, if $\theta_a-\theta_b=\pm\pi$,
then the state in Eq.~\eqref{Path:MIQKD:ConABPhase} becomes
\begin{equation} \label{Path:MIQKD:ConABdeta}
\begin{aligned}
\ket{0110-1001}_{r_0r_1s_0s_1},
\end{aligned}
\end{equation}
which means that either detectors $r_0$ and $s_1$ click, or $r_1$ and $s_0$ click. In all other cases, where $\theta_a-\theta_b=\pm\pi/2$, two random detectors out of four will click, and then Alice and Bob's qubits are independent of each other. Such events will be ruled out by a standard basis-sift procedure. Detection events on only reference (signal) detectors will be ruled out as well, justifying the choice of relevant terms in Eq.~(\ref{Path:MIQKD:ConAB}). In the end, Alice and Bob's bits, determined by relative phases $\theta_a$ and $\theta_b$, will be correlated or anticorrelated conditioned on the detection events in the relay.

Similar to the single-photon case of the original MDI-QKD scheme \cite{Lo:MIQKD:2012}, the key rate formula for our MDI-QKD scheme follows Shor-Preskill's result \cite{ShorPreskill_00,KoashiPreskill_03}
\begin{equation} \label{Path:Sim:Keyrate11}
\begin{aligned}
R \ge Y_{11}[1-f H(e_{11}) -H(e_{11})],
\end{aligned}
\end{equation}
where $Y_{11}$ is the successful detection (trigger in the relay) rate provided that Alice and Bob send out single photons; $e_{11}$ is the quantum bit error rate (QBER); $f$ is the error correction inefficiency (see, e.g, \cite{Brassard:ECeff:1993}; normally, $f\ge1$ with the Shannon limit of $f=1$); and $H(x)$ is the binary entropy function, $H(x)=-x\log_2(x)-(1-x)\log_2(1-x)$. In Appendix \ref{App:Path:SinglePhoton}, we derive the relevant terms in Eq.~\eqref{Path:Sim:Keyrate11} when loss and other nonidealities are taken into account [see Eqs.~\eqref{Path:Model:Y11} and \eqref{Path:Model:e11Y11}].

Our single-photon MDI-QKD scheme offers certain advantages over similar schemes in \cite{Lo:MIQKD:2012} and \cite{Tamaki:MIQKD:2012}. A key difference of our scheme with the original MDI-QKD scheme in \cite{Lo:MIQKD:2012} lies on their encoding procedures. In the scheme of Fig.~\ref{Fig:Path:MIdiag}, the qubit information is encoded in the relative phases of two orthogonal optical modes. The original scheme, on the other hand, relies on polarization encoding, which requires sharing a polarization reference between all three parties and polarization maintenance along the channel. As compared to the MDI-QKD scheme II in \cite{Tamaki:MIQKD:2012}, if used with single photons, both schemes use similar phase encoding, and are resilient to overall phase errors. In our case, the detection setup is simpler: it does not rely on optical switches and, as we will show in the next section, it can operate by using only two detectors. Note that, for the partial BSM part, all schemes require indistinguishable photons, hence filtering before the BSM modules is necessary.

The scheme in Fig.~\ref{Fig:Path:MIdiag} relies on single-photon states for its proper operation. In practice, on-demand single-photon sources can be implemented using parametric down-conversion processes \cite{Shapiro:SPS:07}, or by relying on quasiatomic systems such as quantum dots \cite{ward:on-demand:2005}. In these scenarios, one must consider the effect of multiple photons on system performance, which will be addressed in a separate publication. With recent advancements in compact cost-effective single-photon sources, the reliance on single-photon states in our scheme is not necessarily a setback, especially when considering the simplicity of the BSM module as compared to those proposed in \cite{Lo:MIQKD:2012, Tamaki:MIQKD:2012}. Nevertheless, in Sec.~\ref{Sec:Path:Coherent}, we present the decoy-state version of our protocol, which does not rely on single-photon sources.

The setup in Fig.~\ref{Fig:Path:MIdiag} requires two optical channels for each user, which seems redundant and requires relative phase maintenance between the two channels. In the following section, we show that, by using a simple time-multiplexing trick, one can resolve both issues.

\section{MDI-QKD with time multiplexing} \label{Sec:Path:Time}
\begin{figure}[hbt]
\centering
\resizebox{8cm}{!}{\includegraphics{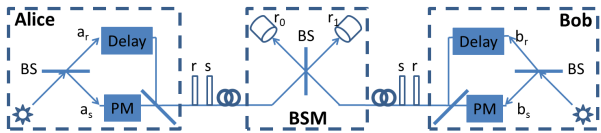}}
\caption{(Color online) A schematic diagram of the time-multiplexed MDI-QKD protocol. Alice and Bob each encodes their qubits onto relative phases of two optical modes separated in time, $a_r$, $a_s$, $b_r$ and $b_s$, respectively. The partial BSM, using a 50:50 BS, is performed in the relay owned by a possibly untrusted party.}
\label{Fig:Path:MITM}
\end{figure}

Instead of path encoding, Alice and Bob can use time multiplexing to separate their reference and signal modes. That can be achieved by using Mach-Zehnder interferometers at the transmitter, as shown in Fig.~\ref{Fig:Path:MITM}. That would result in both reference and signal pulses traveling along the same physical channel. Moreover, if the time delay between the two modes is sufficiently short, we can reliably assume that the relative phase between the reference and signal modes is well preserved along the channel, as  required in Fig.~\ref{Fig:Path:MIdiag}. The BSM module in Fig. \ref{Fig:Path:MITM} is also simpler than that of Fig.~\ref{Fig:Path:MIdiag}, as we are only using two, rather than four, single-photon detectors. It is also simpler than the proposed BSM modules in \cite{Lo:MIQKD:2012, Tamaki:MIQKD:2012}, as it does not require optical switches or phase-to-polarization converters. Similar to any other schemes, time synchronization is required to ensure that the corresponding reference and signal modes will arrive at the right time and properly interfere with each other.

The main problem that must be addressed in this time-multiplexed scheme is the dead time of single-photon detectors. That is, after detection, a detector will be nonresponsive (dead) for a period of time until it resets. The dead time of a detector is caused by the after-pulse effect in avalanche photodiode single-photon detectors. In the time-multiplexed scheme, the detectors are required to detect photons in two consecutive pulses, whose time difference could be short. The dead time of detectors then ultimately limits the repetition rate of the proposed scheme. Here, we propose proper postselection methods to address the dead-time problem.

For the postselection of events, at the BSM module of Fig.~\ref{Fig:Path:MITM}, we consider two scenarios. In the first scenario, we assume that the dead time of single-photon detectors is shorter than the delay in Mach-Zehnder interferometers. In this case, we can use exactly the same postselection technique as described in Sec.~\ref{Sec:Path:MIQKD}. The only difference would be that for the time slot corresponding to signal pulses, detectors $r_0$ and $r_1$ in Fig.~\ref{Fig:Path:MITM} resemble detectors $s_0$ and $s_1$ in Fig.~\ref{Fig:Path:MIdiag}. With recent advances in single-photon detectors with ultrashort dead times \cite{Yuan:Selfdif:2007, shield:dead-time:231113}, one can use a repetition rate as high as 500 MHz with our scheme. In order to go to higher repetition rates, one must use a delay possibly shorter than the detector's dead times. From the discussion of Eq.~\eqref{Path:MIQKD:ConABdetc} and \eqref{Path:MIQKD:ConABdeta}, we notice that only when Alice and Bob's results are correlated (they have used the same phase), is the dead-time issue problematic. In order to resolve this issue, Alice and Bob can further sift out those detection events resulting from the terms in Eq.~\eqref{Path:MIQKD:ConABdetc}. That is, by accepting a factor of $1/2$ loss in the final key rate, we will only keep measurement results in which both $r_0$ and $r_1$ click, each at a different time slot corresponding to the arrival of the reference or signal beams. With the above modified postselection technique, the setup in Fig.~\ref{Fig:Path:MITM} provides comparable secret key generation rates to other single-photon MDI-QKD schemes, while offering a simple and cost-effective structure.

The setup in Fig.~\ref{Fig:Path:MITM} can be easily modified to implement encoding in all three Pauli bases. If we represent the standard basis vectors, i.e., eigenvectors of the $Z$ operator, by a single-photon state in the reference mode and a single-photon state in the signal mode, the encodings implemented by the setup of Fig.~\ref{Fig:Path:MITM} are that of $X$ and $Y$ bases. If one replaces the first beam splitter in the encoder with a polarizing beam splitter, and uses horizontally or vertically polarized light at the source \cite{MXF:MIFluc:2012}, we can use the same setup for $Z$-basis encoding as well. In the case of single-photon sources, which of two bases to choose for the QKD protocol is arbitrary. Once we consider the decoy-state version of our protocol, however, the choice of bases is more crucial. In fact, it turns out that $X$ and $Y$ bases are prone to a larger value of QBER than the $Z$ basis. That is why, in the experimental setup of \cite{Rubenok:MIQKDexp:2012,Liu:MIQKDexp:2012}, $Z$- and $X$-basis encoding is used. A brief analysis of the decoy-state version of the MDI-QKD protocol with $Z$- and $X$-basis encoding is given in \cite{Lo:MIQKD:2012}, which we will rederive within our own setup in Appendix \ref{App:Path:Model}. In the next section, however, we will consider the more challenging $X$- and $Y$-basis encoding for the decoy-state protocol and propose postselection techniques to reduce the QBER in such a scenario.


%

\section{Decoy-state MDI-QKD} \label{Sec:Path:Coherent}
In this section, the decoy-state version of our scheme is presented. A weak laser pulse is perhaps the easiest way to approximate a single-photon state. Due to the multiple-photon component in a laser pulse, in the context of QKD, coherent-state sources are considered to be basis dependent \cite{MXF:EntanglementPDC:2007}. For a basis-dependent source, one can apply decoy-state technique to monitor the channel transmittance of the single-photon component in the source \cite{Hwang:Decoy:2003,Lo:Decoy:2005,MXF:Practical:2005,Wang:Decoy:2005,*Wang:Decoy2:2005}.

In the security proof of the decoy-state scheme, the overall phase of the coherent-state source is assumed to be randomized \cite{Lo:Decoy:2005,LoPreskill:NonRan:2007}. The main problem with using phase-randomized coherent states, in the setup of Fig.~\ref{Fig:Path:MIdiag}, is that the probability of a single photon coming out of Alice's source and a single photon from Bob's source is on the same order as the case of having two photons at Alice's or Bob's, and no photon out of the other source. The former is what we need to generate a secret key bit, whereas the latter could result in random clicks. In fact, the QBER in the scheme of Fig.~\ref{Fig:Path:MIdiag} could be over 20\% if we use a standard decoy-state protocol. In order to resolve this issue, in this section, we assume that Alice and Bob have a common overall phase reference. We then use an improved phase-postselection technique to enhance the efficiency of error correction. We remark that a full security proof of this technique is yet to be addressed.

\subsection{Key rate}


The security analysis for our scheme with decoy states follows from that of \cite{Lo:MIQKD:2012} and \cite{Tamaki:MIQKD:2012}, which rely on the photon-number channel model used in \cite{Lo:Decoy:2005}. The key rate for the original decoy-state QKD is given by \cite{Lo:Decoy:2005, Lo:Vacuum:2005}
\begin{equation} \label{Path:Sim:Keyrate}
\begin{aligned}
R \ge q \{ -Q_{\mu} f H(E_{\mu}) + Q_1[1-H(e_1)] + Q_0 \},
\end{aligned}
\end{equation}
where $q$ is the basis sift factor; the subscript $\mu$ denotes the average number of photons per pulse; $Q_{\mu}$ and $E_{\mu}$ are, respectively, the overall gain and QBER; $Q_1$ and $e_1$ are, respectively, the gain and the error rate of the single-photon components; and $Q_0$ is the gain of the vacuum state (from background).

There are certain details to be considered before applying Eq.~\eqref{Path:Sim:Keyrate} to our case. The basis sift factor $q$ is equal to $1/2$ in the original BB84 protocol due to the fact that half of the time the bases chosen by Alice and Bob disagree. In the efficient BB84 protocol \cite{Lo:EffBB84:2005}, however, the factor $q$ can approach 1 in the infinite-size key limit. In our MDI-QKD scheme, there is an extra factor of $1/2$ due to the partial BSM described in Sec.~\ref{Sec:Path:MIQKD}. If one uses the correlation sift technique to handle the dead time problem as discussed in Sec.~\ref{Sec:Path:Time}, another factor of $1/2$ must also be accounted for. In our key-rate analysis, described in Appendix \ref{App:Path:Model}, we merge this factor, $q$, into other gain factors to obtain
\begin{equation} \label{Path:Sim:KeyrateMI}
\begin{aligned}
R &\ge   Q_{11}[1-H(e_{11})] + Q'_{0\mu_b} - I_{ec} ,\\
I_{ec} &= Q_{\mu_a\mu_b} f H(E_{\mu_a\mu_b}),
\end{aligned}
\end{equation}
where $I_{ec}$ is the cost of error correction; $Q_{\mu_a\mu_b}$ ($E_{\mu_a\mu_b}$) is the overall gain (QBER) when Alice and Bob, respectively, use an average photon number of $\mu_a$ and $\mu_b$; $Q_{11}$ ($e_{11}$) is the gain (QBER) when both sources generate single-photon states; and $Q'_{0\mu_b} = \exp(-\mu_a) Q_{0\mu_b}$ is the probability that there is no photon from Alice's side {\em and} a successful BSM occurs. Appendix \ref{App:Path:Model} provides detailed definitions for the above parameters.

Here, we assume that Alice and Bob use forward classical communication (Alice to Bob) for error correction and privacy amplification, which leads to the $Q'_{0\mu_b}$ term in the key rate formula in Eq.~\eqref{Path:Sim:KeyrateMI}. The security argument behind it is that when Alice sends out vacuum states, Eve gains nothing about Alice's qubits by measuring the state in the channel \cite{Lo:Vacuum:2005,Koashi:Comp:09}. Of course, one can assume that Alice and Bob perform reverse reconciliation, in which case $Q'_{0\mu_b}$ must be replaced with $Q'_{\mu_a 0} = \exp(-\mu_b)Q_{\mu_a 0}$. We emphasize that, because of relying on two photons for a successful BSM, $Q'_{0\mu_b}$ is on the same order of $Q_{\mu_a\mu_b}$ for coherent-state sources. Thus, its contribution is significant. On the contrary, for the key rate of a regular decoy-state QKD given by Eq.~\eqref{Path:Sim:Keyrate}, the contribution from the vacuum state, $Q_0$, is insignificant due to the fact that normally the background count rate is much lower than $Q_\mu$.


Note that, in MDI-QKD schemes, postprocessing can be performed separately on bit strings obtained from different bases \cite{MXF:Finite:2011}, or from different detection events (correlated and anticorrelated). In practice, this extra information may be useful for error correction \cite{MXF:DDIQKD:2012}.

\subsection{MDI-QKD with phase postselection} \label{Sub:Path:PhasePP}
\begin{figure}[hbt]
\centering
\resizebox{8cm}{!}{\includegraphics{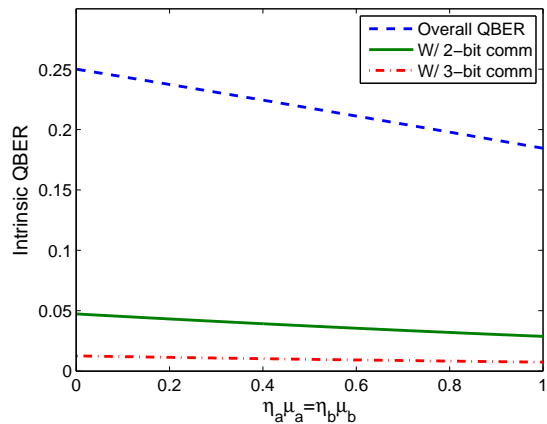}}
\caption{(Color online) QBER for the MDI-QKD scheme in Fig.~\ref{Fig:Path:MIdiag} with coherent-state sources, conditioned on partial knowledge of the overall phase. The overall QBER, calculated by Eq.~\eqref{Path:Model:EQmumuIntrInt}, represents the case where no phase information is shared. Curves labeled ``W/ 2(3)-bit comm" represent cases where Alice and Bob postselect states coming from the same phase region, out of $N=4 (8)$ phase bands in Eq.~\eqref{Path:Model:PhasePartition}. The conditional QBER is calculated numerically using Eq.~\eqref{Path:Model:EQmuIntrps}. No background noise or misalignment is assumed.}
\label{Fig:Path:QBER}
\end{figure}

As mentioned before, the intrinsic QBER in our decoy-state scheme can be very high if one uses a fully phase-randomized coherent source. Note that randomization of the overall phase over $[0,2 \pi)$ can be regarded as randomization over one of the following $N$ regions:
\begin{equation} \label{Path:Model:PhasePartition}
\begin{aligned}
\left\{[\frac{m\pi}{N},\frac{(m+1)\pi}{N})\cup[\frac{(m+N)\pi}{N},\frac{(m+N+1)\pi}{N}) \mid m=0,1,2,N-1\right\}.
\end{aligned}
\end{equation}
The choice of the region and the overall phase therein are random at the source. In the conventional decoy-state protocol, no information about the overall phase is exchanged between two users. In our scheme, before error correction, Alice and Bob would reveal which region they had used. They would only keep raw key bits for which they both have used the same phase region. This extra information would reduce the cost of error correction, because the QBER is different for the raw key from different regions. In fact, in Fig.~\ref{Fig:Path:QBER}, we can see that by using $N=4$ and $N=8$, corresponding to, respectively, two and three bits of classical information, the QBER has been reduced to below 5\% and 1.3\%. In Fig.~\ref{Fig:Path:QBER}, we have assumed that $\eta_a\mu_a=\eta_b\mu_b$, where $\eta_a$ and $\eta_b$ are, respectively, the total transmission efficiency for Alice and Bob's paths. In Appendix \ref{App:Path:Model}, we show that the QBER $E_{\mu_a\mu_b}$ is minimized under this condition. Note that Alice and Bob do not need to control the phases of coherent states precisely, which is practically challenging. Instead, as long as they know the phase partition in Eq.~\eqref{Path:Model:PhasePartition} with a high probability, the phase postselection method proposed here can be implemented. We leave the case where Alice and Bob do not exactly know the phase values for future study.

Conditioned on the classical bits Alice sends to Bob, the cost of error correction in Eq.~\eqref{Path:Sim:KeyrateMI} is given by
\begin{equation} \label{Path:Sim:IecAdv}
\begin{aligned}
I_{ec} &= \sum_{m}Q^{m} f H(E^{m}),
\end{aligned}
\end{equation}
where $m$ is the partition index in Eq.~\eqref{Path:Model:PhasePartition}, and $Q^m$ and $E^m$ are the corresponding conditional gain and QBER [see Eqs.~\eqref{Path:Model:QmuIntrps} and \eqref{Path:Model:EQmuIntrps}].

Our numerical calculations show that the key rate given by Eqs.~\eqref{Path:Sim:KeyrateMI} and \eqref{Path:Sim:IecAdv} is not positive for the parameter set listed in Table \ref{Tab:Path:SimPara}. If, however, one assumes the gain and error rates of single-photon states are evenly distributed over the partitions of Eq.~\eqref{Path:Model:PhasePartition}, the key rate formula, Eq.~\eqref{Path:Sim:KeyrateMI}, becomes
\begin{equation} \label{Path:Sim:Keyrate3bit}
\begin{aligned}
R &\ge   \frac{1}{N}Q_{11}[1-H(e_{11})] - Q^{m} f H(E^{m})|_{m=0}, \\
\end{aligned}
\end{equation}
where we take the lower bound of $Q_{0\mu_b}'\ge0$. Here, we only keep the term corresponding to $m=0$ in Eq.~\eqref{Path:Sim:IecAdv}, in which case $E^m$ is minimized.

\subsection{Key-rate comparison}
In this section, we numerically compare the secret key generation rate for the MDI-QKD schemes proposed here using single-photon and decoy-coherent states with that of \cite{Lo:MIQKD:2012}.
For fair comparison, we use the same parameter values used in \cite{Lo:MIQKD:2012} for our numerical evaluation, which follow the experiment reported in \cite{Ursin:PDC144:2007} (see also \cite{MXF:TriggeringPDC:2008}). The numerical parameters used are listed in Table \ref{Tab:Path:SimPara}. We have used Eq.~\eqref{Path:Sim:Keyrate3bit} and formulas in Appendixes \ref{App:Path:SinglePhoton} and \ref{App:Path:Model} to evaluate the key rate of our decoy-state scheme.
\begin{table}[hbt]
\centering
\begin{tabular}{ccccccc}
\hline
Quantum efficiency & $p_d$ & $f$ & $e_d$ \\
\hline
14.5\% & $3.0\times 10^{-6}$ & 1.16 & 1.5\% \\
\hline
\end{tabular}
\caption{List of experimental parameters used in numerical results: $p_d$ is the background count rate per detector; $f$ is the error correction inefficiency; and $e_d$ is the misalignment error between Alice and Bob, which characterizes the stability of the relative phases at the encoders and through the channel. Note that two detectors are used in the original experiment \cite{Ursin:PDC144:2007}, thus, $p_d$ should be roughly half of the total background count rate.} \label{Tab:Path:SimPara}
\end{table}

\begin{figure}[hbt]
\centering
\resizebox{8cm}{!}{\includegraphics{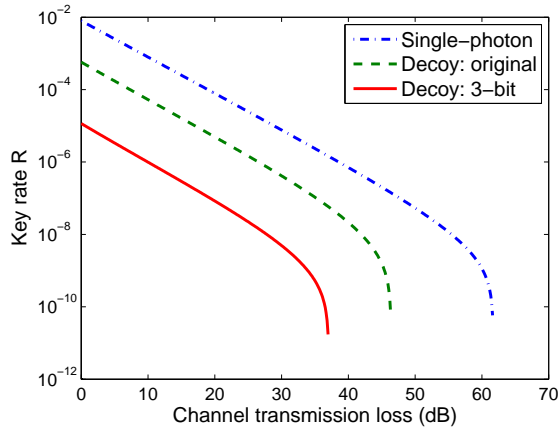}}
\caption{(Color online) Key rate comparison for single-photon and decoy-state MDI-QKD schemes. The setup parameters are listed in Table \ref{Tab:Path:SimPara}. The solid line indicates the key rate for the $X$-$Y$-basis-encoding scheme with decoy states plus three-bit communication for overall phase postselection, shown in Eq.~\eqref{Path:Sim:Keyrate3bit}. The dashed line shows the performance of the original MDI-QKD with $X$-$Z$-basis encoding. The related formulas for simulation can be found in Appendixes \ref{App:Path:SinglePhoton} and \ref{App:Path:Model}. The $\mu$'s are optimized for the two decoy-state curves.}
\label{Fig:Path:Keyrate}
\end{figure}

Figure \ref{Fig:Path:Keyrate} shows the secret key generation rate for the three schemes mentioned above. The middle curve corresponds to that of the original MDI-QKD scheme \cite{Lo:MIQKD:2012}, obtained from Eq.~\eqref{Rate:Decoy:Orig}. It can be seen that while our single-photon scheme can outperform the original MDI-QKD scheme, our decoy-state protocol falls short of achieving the same performance. Nevertheless, both our schemes offer simpler BSM modules than what proposed in \cite{Lo:MIQKD:2012} and \cite{Tamaki:MIQKD:2012}. One of the reasons why the key rate of our scheme is lower than the original scheme is the additional phase postselection factor of $N=8$.

\begin{figure}[hbt]
\centering
\resizebox{8cm}{!}{\includegraphics{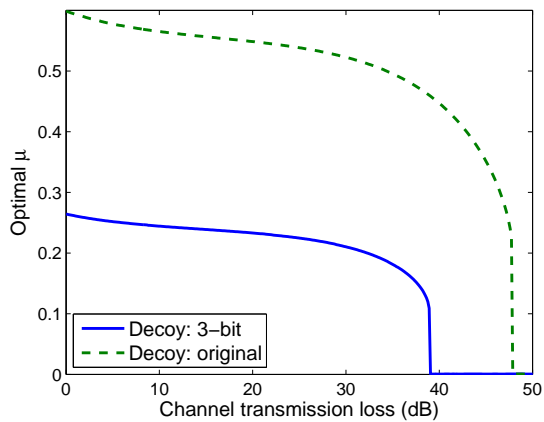}}
\caption{(Color online) Optimal average photon number of coherent state for the $X$-$Y$-basis-encoding scheme (solid line) and the original MDI-QKD scheme (dashed line, $X$-$Z$ encoding) with decoy states. The setup parameters are listed in Table \ref{Tab:Path:SimPara} and the related formulas can be found in Appendix \ref{App:Path:Model}.}
\label{Fig:Path:Optmu}
\end{figure}

The key rate in our scheme (and the original scheme \cite{Lo:MIQKD:2012}) scales better with distance than that of scheme I in \cite{Tamaki:MIQKD:2012}. The latter yields a key rate scaling quadratically with the channel transmittance, whereas in our scheme, it scales linearly. This is because the optimal $\mu$ of scheme I is on the same order of the transmission efficiency as shown in \cite{Tamaki:MIQKD:2012}, whereas the optimal $\mu$ in our scheme is on the order of 1, as shown in Fig.~\ref{Fig:Path:Optmu}.

From Fig.~\ref{Fig:Path:Optmu}, one can see that the optimal $\mu$ in our scheme is smaller than the one given by the original MDI-QKD. This is because in our scheme, multiphoton states would introduce false triggers in the BSM relay, which causes an error rate of 1/2. In order to reduce such an effect, a smaller $\mu$ should be used. This is another reason why the key rate of our decoy-state scheme is lower than the original one, as shown in Fig.~\ref{Fig:Path:Keyrate}. We remark that the key rate is quite stable with certain changes of $\mu$, except for the regime where the channel loss is close to the maximal tolerable one.





%

\section{Concluding remarks} \label{Sec:Path:Conclusion}
Measurement-device-independent schemes have been proposed to close the detection loopholes in QKD systems. In this paper, we presented phase-encoding MDI-QKD setups that could offer certain practical advantages over previously proposed schemes in \cite{Lo:MIQKD:2012, Tamaki:MIQKD:2012}. If implemented with single-photon states, our scheme enjoys a simple detection setup, consisting of only two single-photon detectors and a 50:50 beam splitter, with little or no compromise on the performance.
Polarization or overall phase maintenance through quantum channels is not required in our schemes, either. This is an advantage over the polarization scheme in \cite{Lo:MIQKD:2012} or the phase-encoding schemes in \cite{Tamaki:MIQKD:2012}. There are different decoy-state versions one can implement using our setup. The original MDI-QKD is effectively using an $X$-$Z$-basis encoding, with a lower QBER for the $Z$ basis. Here we showed that, by a proper overall phase selection scheme, we could achieve positive secret key rates even if we used $X$ and $Y$ bases for encoding over a moderately long range of distances. We remark that the security of the phase postselection technique described in Appendix \ref{App:Path:PhasePost} needs to be further investigated. Similar questions were raised for DIQKD schemes \cite{MXF:DDIQKD:2012}.

For a full key rate analysis, finite-key effects and statistical fluctuations must also be considered. That would include fluctuation analysis for decoy states \cite{MXF:Practical:2005} as well as phase error estimations \cite{MXF:Finite:2011,Fung:Finite:2010}. We remark that the statistical fluctuation analysis for the MDI-QKD with decoy states has been recently presented \cite{MXF:MIFluc:2012}. Other finite-key effects, such as authentication, are expected to be negligible compared to the above two effects in a large parameter set \cite{MXF:Finite:2011}.

In the current MDI-QKD realizations \cite{Lo:MIQKD:2012,Tamaki:MIQKD:2012}, including the one we propose here, we assume Alice and Bob use the same source settings. It is interesting to study the case where two source settings are different. For instance, one of the parties uses coherent states with the decoy-state protocol and the other one uses single-photon states.

We finally remark that our proposed scheme can be easily adapted to quantum network settings \cite{Kimble_QInternet_Nature, Razavi_MA_QKD}. For a single-hop network, the switching center is a collection of several BSM modules along with switching and controlling devices. For longer distances, one can connect two switching centers with quantum repeaters \cite{DLCZ_01, Razavi.DLCZ.06, Razavi.Amirloo.10, Sangouard:single-photon:2007, Razavi.Lutkenhaus.09, NLoPiparo:Qrep:2012} and effectively enable any two users to exchange secret keys. One of the key advantages of such a setup is the simplicity and the low cost of local users' equipment (the optical source), while the more expensive part, i.e., detectors, are shared among all users.

\subsection*{Acknowledgments}
The authors would like to thank T.-Y.~Chen, N.~Jain, H.-K.~Lo, B.~Qi, K.~Tamaki, W.~Tittel, F.~Xu and Q.~Zhang for enlightening discussions. The authors gratefully acknowledge the financial support from National Basic Research Program of China Grants No.~2011CBA00300 and No.~2011CBA00301, National Natural Science Foundation of China Grants No.~61073174, No.~61033001, and No.~61061130540, the 1000 Youth Fellowship program in China, the European Community's Seventh Framework Programme under Grant Agreement 277110, the UK Engineering and Physical Science Research Council Grant No.~EP/J005762/1, NSERC, CRC program, CIFAR, and QuantumWorks.

\begin{appendix}

\section{MDI-QKD with single-photon states} \label{App:Path:SinglePhoton}
In this appendix, we will consider channel losses, background counts, and misalignment errors for the scheme introduced in Sec.~\ref{Sec:Path:MIQKD}.

The initial joint state of Alice and Bob is given by Eq.~\eqref{Path:MIQKD:OriAB}. After passing through lossy channels, modeled by beam splitters with transmissivities $\eta_a$ and $\eta_b$, and considering the normalization factors, the state in Eq.~\eqref{Path:MIQKD:OriAB} becomes a mixed state as follows:
\begin{equation} \label{initial_state}
\frac{\eta_a \eta_b}{4} \ket{\psi_{11}} \bra{\psi_{11}} + \frac{\eta_a (1-\eta_b)}{2} \ket{\psi_{10}} \bra{\psi_{10}}
+ \frac{(1-\eta_a) \eta_b}{2} \ket{\psi_{01}} \bra{\psi_{01}} + (1-\eta_a)(1- \eta_b) \ket{\psi_{00}} \bra{\psi_{00}},
\end{equation}
where
\begin{equation} \label{Path:Model:OriJointAB11}
\begin{aligned}
\ket{\psi_{11}} &= \ket{1010}+e^{i\theta_a}\ket{0110}+e^{i\theta_b}\ket{1001} + e^{i(\theta_a+\theta_b)}\ket{0101}, \\
\ket{\psi_{10}} &= \ket{1000}+e^{i\theta_a}\ket{0100}, \\
\ket{\psi_{01}} &= \ket{0010}+e^{i\theta_b}\ket{0001}, \\
\ket{\psi_{00}} &= \ket{0000}. \\
\end{aligned}
\end{equation}
Here, as a shorthand notation, in the above equation $\ket{abcd}$ represents the joint number state $\ket{abcd}_{a_r a_s b_r b_s}$.

The state in Eq.~\eqref{initial_state} will then pass through beam splitters in the relay, as described in Eq.~\eqref{Path:MIQKD:DetTrans}. The state $\ket{\psi_{11}}$ is transformed to
\begin{equation} \label{Path:Model:11relay1st}
\begin{aligned}
&\left[\ket{1010}+e^{i\theta_a}\ket{0110}+e^{i\theta_b}\ket{1001} +e^{i(\theta_a+\theta_b)}\ket{0101}\right] \\
&\mapsto \frac{1}{\sqrt{2}}\ket{02-20}\ket{00} +\frac{1}{2}e^{i\theta_a}\ket{01-10}\ket{01+10} \\ &\;\;\;+\frac{1}{2}e^{i\theta_b}\ket{01+10}\ket{01-10} +\frac{1}{\sqrt{2}}e^{i(\theta_a+\theta_b)}\ket{00}\ket{02-20}, \\
\end{aligned}
\end{equation}
where the optical modes $a_r$, $a_s$, $b_r$, and $b_s$ are mapped to $r_0$, $r_1$, $s_0$, and $s_1$. Here, we used the following transformations corresponding to a 50:50 beam splitter
\begin{equation} \label{Path:Model:DetTransadd}
\begin{aligned}
\ket{0}_{a_r}\ket{0}_{b_r} &\mapsto \ket{0}_{r_0}\ket{0}_{r_1}, \\
\ket{1}_{a_r}\ket{1}_{b_r} &\mapsto [\ket{0}_{r_0}\ket{2}_{r_1} - \ket{2}_{r_0}\ket{0}_{r_1}]/\sqrt{2}, \\
\ket{0}_{a_s}\ket{0}_{b_s} &\mapsto \ket{0}_{s_0}\ket{0}_{s_1}, \\
\ket{1}_{a_s}\ket{1}_{b_s} &\mapsto [\ket{0}_{s_0}\ket{2}_{s_1} - \ket{2}_{s_0}\ket{0}_{s_1}]/\sqrt{2}, \\
\end{aligned}
\end{equation}
where the arrived photons are assumed to be indistinguishable, say, by passing through proper filters, such as polarization and frequency filters, before the partial BSM.

Similarly, other states in Eq.~\eqref{Path:Model:OriJointAB11} are transformed to
\begin{equation} \label{Path:Model:11relay234}
\begin{aligned}
\ket{1000}+e^{i\theta_a}\ket{0100} &\mapsto [\ket{01+10}\ket{00}+e^{i\theta_a}\ket{00}\ket{01+10}]/\sqrt{2}, \\
\ket{0010}+e^{i\theta_b}\ket{0001} &\mapsto [\ket{01-10}\ket{00}+e^{i\theta_a}\ket{00}\ket{01-10}]/\sqrt{2}, \\
\ket{0000} &\mapsto \ket{0000}. \\
\end{aligned}
\end{equation}

Define a successful partial BSM event to be the case when exactly one of the two detectors in each mode of the relay (that is, $r_0$ and $s_0$, $r_0$ and $s_1$, $r_1$ and $s_0$, or $r_1$ and $s_1$) clicks, as shown in Fig.~\ref{Fig:Path:MIdiag}.
The yield, $Y_{11}$, is defined as the probability to have a successful measurement event, given that both Alice and Bob send out single-photon states and choose the same basis (that is, $\theta_a-\theta_b=0,\pi$).

When Alice and Bob's bits are correlated ($\theta_a-\theta_b=0$),  Eq.~\eqref{Path:Model:11relay1st} becomes
\begin{equation} \label{Path:Model:11relay1st0}
\begin{aligned}
&\left[\ket{1010}+e^{i\theta_a}\ket{0110}+e^{i\theta_b}\ket{1001} +e^{i(\theta_a+\theta_b)}\ket{0101}\right] \\
&\mapsto \frac{1}{\sqrt{2}}\ket{02-20}\ket{00} +e^{i\theta_a}\ket{0101-1010} +\frac{1}{\sqrt{2}}e^{2i\theta_a}\ket{00}\ket{02-20}, \\
\end{aligned}
\end{equation}
where the second term on the right hand side is the postselected term mentioned in Eq.~\eqref{Path:MIQKD:ConABdetc}.

With Eqs.~\eqref{initial_state}, \eqref{Path:Model:OriJointAB11}, \eqref{Path:Model:11relay234}, and \eqref{Path:Model:11relay1st0}, we can calculate the probability for a single click in each mode when $\theta_a-\theta_b=0$,
\begin{equation} \label{Path:Model:Y114cases0}
\begin{aligned}
{}_0Y^{r_0s_0}_{11} &= {}_0Y^{r_1s_1}_{11} = (1-p_d)^2\left[ \frac{\eta_a\eta_b}{4} +\left(\frac{\eta_a+\eta_b}{2}-\frac{3\eta_a\eta_b}{4}\right)p_d +(1-\eta_a)(1-\eta_b)p_d^2 \right], \\
{}_0Y^{r_0s_1}_{11} &= {}_0Y^{r_1s_0}_{11} = (1-p_d)^2\left[ \left(\frac{\eta_a+\eta_b}{2}-\frac{3\eta_a\eta_b}{4}\right)p_d +(1-\eta_a)(1-\eta_b)p_d^2 \right], \\
\end{aligned}
\end{equation}
where $p_d$ is the background rate for one detector ($p_d\approx Y_0/2$). Due to the symmetry, the probabilities for the case when $\theta_a-\theta_b=\pi$ are similar:
\begin{equation} \label{Path:Model:Y114casespi}
\begin{aligned}
{}_{\pi}Y^{r_0s_0}_{11} &= {}_{\pi}Y^{r_1s_1}_{11} = {}_{0}Y^{r_0s_1}_{11}, \\ {}_{\pi}Y^{r_0s_1}_{11} &= {}_{\pi}Y^{r_1s_0}_{11} = {}_{0}Y^{r_0s_0}_{11}. \\
\end{aligned}
\end{equation}
Thus the yield $Y_{11}$, defined as the total probability to have a successful measurement event when Alice and Bob use the same basis, is given by the summation of the terms in Eq.~\eqref{Path:Model:Y114cases0} [or Eq.~\eqref{Path:Model:Y114casespi}] as follows:
\begin{equation} \label{Path:Model:Y11}
\begin{aligned}
Y_{11} &= {}_{0}Y^{r_0s_0}_{11} + {}_{0}Y^{r_1s_1}_{11} + {}_{0}Y^{r_0s_1}_{11} + {}_{0}Y^{r_1s_0}_{11} \\
&=(1-p_d)^2\left[ \frac{\eta_a\eta_b}{2} +(2\eta_a+2\eta_b-3\eta_a\eta_b)p_d +4(1-\eta_a)(1-\eta_b)p_d^2 \right]. \\
\end{aligned}
\end{equation}
When $p_d=0$, $Y_{11}=\eta_a\eta_b/2$, which is reasonable because the probability of the two optical modes each containing exactly one photon is $1/2$.

An error may occur when $\theta_a-\theta_b=0$ but an anticorrelated detection signal comes out; that is, detectors $r_0$ and $s_1$, or $r_1$ and $s_0$ click. Thus, the error rate due to background noise is given by
\begin{equation} \label{Path:Model:e11Y0}
\begin{aligned}
e'_{11}Y_{11} &= {}_{0}Y^{r_0s_1}_{11} + {}_{0}Y^{r_1s_0}_{11} \\
&=(1-p_d)^2\left[ \left(\eta_a+\eta_b-\frac{3\eta_a\eta_b}{2}\right)p_d +2(1-\eta_a)(1-\eta_b)p_d^2 \right] \\
&= e_0(1-p_d)^2\left[ (2\eta_a+2\eta_b-3\eta_a\eta_b)p_d +4(1-\eta_a)(1-\eta_b)p_d^2 \right] , \\
\end{aligned}
\end{equation}
where $e_0=1/2$ is the error rate of a random (background) noise. Now considering possible phase errors, i.e., the deviation of $\Delta_\theta = \theta_a - \theta_b$ from its nominal value, the total error rate is given by
\begin{equation} \label{Path:Model:e11Y11}
\begin{aligned}
e_{11}Y_{11} &= (1-p_d)^2\left[ e_d\frac{\eta_a\eta_b}{2} +e_0(2\eta_a+2\eta_b-3\eta_a\eta_b)p_d +4e_0(1-\eta_a)(1-\eta_b)p_d^2 \right] \\
&= e_0Y_{11}-(e_0-e_d)(1-p_d)^2\frac{\eta_a\eta_b}{2}, \\
\end{aligned}
\end{equation}
where $e_d$ is approximately the variance of $\Delta_\theta$, accounting for channel relative-phase distortions (misalignment).

\section{MDI-QKD with Decoy States} \label{App:Path:Model}
In this Appendix, we calculate the key parameters in Eq.~\eqref{Path:Sim:KeyrateMI}. As pointed out in \cite{Lo:MIQKD:2012}, with an infinite number of decoy states, theses parameters can be accurately estimated. One of the key assumptions in decoy-state analysis is the phase randomization at the source \cite{MXF:Practical:2005}. According to the photon channel model \cite{Lo:Decoy:2005}, with phase randomization, a coherent state can be regarded as a mixture of Fock states. In fact, any state can be treated as a mixture of Fock states when the phase of the Fock state component is randomized. In this appendix, we first consider the case when the phase is randomized over $[0, 2\pi)$, and then consider our phase postselection technique. For simplicity, we only consider the limit of the efficient BB84 scheme, where the basis-sift factor is approaching 1  \cite{Lo:EffBB84:2005}. That is, the difference between relative phases set by Alice and Bob in Fig.~\ref{Fig:Path:MIdiag} is either 0 or $\pi$ almost surely.




\subsection{Coherent states with full phase randomization}
Now, let us consider the case where phase-randomized coherent states are used \footnote{Any other states can be applied here with certain modifications in the formulas, mainly in the photon number distribution.}. According to the Poisson distribution of photon numbers in a coherent state, the gain of single-photon states $Q_{11}$ defined as the probability that both Alice and Bob send out single-photon states with the same basis \emph{and} obtain a successful partial BSM is given by
\begin{equation} \label{Path:Model:SourceP11}
\begin{aligned}
Q_{11} = \mu_a\mu_b e^{-\mu_a-\mu_b}Y_{11}, \\
\end{aligned}
\end{equation}
where the yield $Y_{11}$ is given by Eq.~\eqref{Path:Model:Y11}.

Next, we evaluate the overall gain and QBER. Alice and Bob prepare coherent states with intensities $\mu_a$ and $\mu_b$, respectively, and randomize the phases
\begin{equation} \label{Path:Model:OriAB}
\begin{aligned}
\ket{e^{i\phi_a}\sqrt{\mu_a}}_{a}\ket{e^{i\phi_b}\sqrt{\mu_b}}_{b},
\end{aligned}
\end{equation}
where $\phi_a$ and $\phi_b$ are the overall randomized phases. Then, the photon sources are split into two orthogonal optical modes, labeled by $r$ and $s$, by 50:50 beam splitters, as described in Sec.~\ref{Sec:Path:MIQKD},
\begin{equation} \label{Path:Model:RelayAB}
\begin{aligned}
\ket{e^{i\phi_a}\sqrt{\frac{\mu_a}{2}}}_{a_r} \ket{e^{i(\theta_a+\phi_a)}\sqrt{\frac{\mu_a}{2}}}_{a_s}
\ket{e^{i\phi_b}\sqrt{\frac{\mu_b}{2}}}_{b_r} \ket{e^{i(\theta_b+\phi_b)}\sqrt{\frac{\mu_b}{2}}}_{b_s}, \\
\end{aligned}
\end{equation}
where $\theta_a$ and $\theta_b$ are the relative phases Alice and Bob want to encode, as shown in Fig.~\ref{Fig:Path:MIdiag}.
Transmitting through lossy channels, modeled by beam splitters, the joint state arrived at the relay can be expressed by
\begin{equation} \label{Path:Model:RelayAB2}
\begin{aligned}
\ket{e^{i\phi_a}\sqrt{\frac{\eta_a\mu_a}{2}}}_{a_r} \ket{e^{i(\theta_a+\phi_a)}\sqrt{\frac{\eta_a\mu_a}{2}}}_{a_s}
\ket{e^{i\phi_a}\sqrt{\frac{\eta_b\mu_b}{2}}}_{b_r} \ket{e^{i(\theta_b+\phi_b)}\sqrt{\frac{\eta_b\mu_b}{2}}}_{b_s}. \\
\end{aligned}
\end{equation}
After passing through the beam splitters in the relay, the state is transformed into, according to Eq.~\eqref{Path:MIQKD:DetTrans}, four detection modes, $r_0$, $r_1$, $s_0$ and $s_1$,
\begin{equation} \label{Path:Model:AB4mode}
\begin{aligned}
&\ket{e^{i\phi_a}\frac{\sqrt{\eta_a\mu_a}}{2}+e^{i\phi_b}\frac{\sqrt{\eta_b\mu_b}}{2}}_{r_0} \ket{e^{i\phi_a}\frac{\sqrt{\eta_a\mu_a}}{2}-e^{i\phi_b}\frac{\sqrt{\eta_b\mu_b}}{2}}_{r_1} \\
&\otimes
\ket{e^{i(\theta_a+\phi_a)}\frac{\sqrt{\eta_a\mu_a}}{2}+e^{i(\theta_b+\phi_b)}\frac{\sqrt{\eta_b\mu_b}}{2}}_{s_0}
\ket{e^{i(\theta_a+\phi_a)}\frac{\sqrt{\eta_a\mu_a}}{2}-e^{i(\theta_b+\phi_b)}\frac{\sqrt{\eta_b\mu_b}}{2}}_{s_1}. \\
\end{aligned}
\end{equation}
Therefore, the detection probabilities for the four detectors are given by
\begin{equation} \label{Path:Model:Prob4mode}
\begin{aligned}
D_{r_0} &= 1-(1-p_d)\exp\left(- \left|e^{i\phi_a}\frac{\sqrt{\eta_a\mu_a}}{2}+e^{i\phi_b}\frac{\sqrt{\eta_b\mu_b}}{2}\right|^2 \right), \\
D_{r_1} &= 1-(1-p_d)\exp\left(- \left|e^{i\phi_a}\frac{\sqrt{\eta_a\mu_a}}{2}-e^{i\phi_b}\frac{\sqrt{\eta_b\mu_b}}{2}\right|^2 \right), \\
D_{s_0} &= 1-(1-p_d)\exp\left(- \left|e^{i(\theta_a+\phi_a)}\frac{\sqrt{\eta_a\mu_a}}{2}+e^{i(\theta_b+\phi_b)}\frac{\sqrt{\eta_b\mu_b}}{2}\right|^2 \right), \\
D_{s_1} &= 1-(1-p_d)\exp\left(- \left|e^{i(\theta_a+\phi_a)}\frac{\sqrt{\eta_a\mu_a}}{2}-e^{i(\theta_b+\phi_b)}\frac{\sqrt{\eta_b\mu_b}}{2}\right|^2 \right). \\
\end{aligned}
\end{equation}
For simplicity, we use the following notations:
\begin{equation} \label{Path:Model:CoherentNotations}
\begin{aligned}
\mu' &= \eta_a\mu_a+\eta_b\mu_b, \\
\Delta_{\phi} &= \phi_b-\phi_a, \\
x &= \sqrt{\eta_a\mu_a\eta_b\mu_b}/2, \\
y &= (1-p_d)e^{-{\mu'}/4}. \\
\end{aligned}
\end{equation}
Here, $\mu'$ denotes the average number of photons reaching the relay, and $\Delta_{\phi}$ denotes the difference between the random overall phases set by Alice and Bob, which should be integrated over $[0,2\pi)$. Then, Eq.~\eqref{Path:Model:Prob4mode} can be simplified to
\begin{equation} \label{Path:Model:Prob4modeSimpl}
\begin{aligned}
D_{r_0} &= 1-ye^{-x\cos\Delta_{\phi}}, \\
D_{r_1} &= 1-ye^{x\cos\Delta_{\phi}}, \\
D_{s_0} &= 1-ye^{-x\cos(\Delta_{\phi}+\theta_a-\theta_b)}, \\
D_{s_1} &= 1-ye^{x\cos(\Delta_{\phi}+\theta_a-\theta_b)}. \\
\end{aligned}
\end{equation}

The gain $Q_{\mu_a\mu_b}$ is defined as the probability that Alice and Bob choose the same basis \emph{and} obtain a successful measurement, and is given by
\begin{equation} \label{Path:Model:SuccProb}
\begin{aligned}
Q_{\mu_a\mu_b} &= [D_{r_0}(1-D_{r_1})+(1-D_{r_0})D_{r_1}] [D_{s_0}(1-D_{s_1})+(1-D_{s_0})D_{s_1}]. \\
\end{aligned}
\end{equation}
Strictly speaking, Eq.~\eqref{Path:Model:SuccProb} should be averaged over random phases $\phi_a$ and $\phi_b$, and different values for $\theta_a$ and $\theta_b$. We delay this averaging until the last stage.
By substituting  Eq.~\eqref{Path:Model:Prob4modeSimpl} into Eq.~\eqref{Path:Model:SuccProb}, we have
\begin{equation} \label{Path:Model:QmumuGen}
\begin{aligned}
Q_{\mu_a\mu_b} &= y^2(e^{-x\cos\Delta_{\phi}}+e^{x\cos\Delta_{\phi}}-2y)^2, \\
\end{aligned}
\end{equation}
where we use the fact that $|\theta_a-\theta_b|=0,\pi$ when Alice and Bob choose the same basis. For a small $\mu'$ (thus, $\sqrt{\eta_a\mu_a\eta_b\mu_b}\le\mu'/2$ is also small) and $p_d=0$, the gain, Eq.~\eqref{Path:Model:QmumuGen}, will be approximated by
\begin{equation} \label{Path:Model:QmumuApprox}
\begin{aligned}
Q_{\mu_a\mu_b} &\rightarrow \left(\frac{\mu'}{2}\right)^2. \\
\end{aligned}
\end{equation}
Note that Eq.~\eqref{Path:Model:QmumuApprox} is independent of $\Delta_{\phi}$, which can be understood as follows. In the weak coherent-state limit ($\eta_a\mu_a\approx\eta_b\mu_b\ll1$), there are two dominant terms in the relay: single-photon states on both sides versus a vacuum state on one arm and a two-photon state on the other. The vacuum state is not affected by the phase shift, and then the phase of the two-photon state will behave like an overall phase, which does not affect the measurement result. Also, as shown in Sec.~\ref{Sec:Path:MIQKD}, the randomized phase does not affect the partial BSM of single-photon states. Thus, Eq.~\eqref{Path:Model:QmumuApprox} is independent of $\Delta_{\phi}$.

Now, we take the integral of $\Delta_{\phi}$ for Eq.~\eqref{Path:Model:QmumuGen},
\begin{equation} \label{Path:Model:QmumuInt}
\begin{aligned}
Q_{\mu_a\mu_b} &= 2y^2[1+2y^2-4yI_0(x)+I_0(2x)], \\
\end{aligned}
\end{equation}
where $I_0(x)$ is the modified Bessel function of the first kind. For small values of $x$, one can take the first-order approximation to $I_0(x) \approx 1+x^2/4$ to verify that Eq.~\eqref{Path:Model:QmumuInt} approaches Eq.~\eqref{Path:Model:QmumuApprox}, when $p_d=0$ and $\mu'$ is small. Note that in this weak coherent-state limit, the overall gain $Q_{\mu_a\mu_b}$ \emph{cannot} be approximated by the gain of single-photon states, $Q_{11}$ in Eq.~\eqref{Path:Model:SourceP11}, because two-photon states cannot be neglected in this case. This is different from regular decoy-state QKD \cite{MXF:Practical:2005}, where a coherent state can be approximated as a single-photon state when the intensity is low enough. We remark that this property will make the statistical fluctuation analysis more complicated for MDI-QKD.

Using Eq.~\eqref{Path:Model:QmumuInt}, we calculate $Q'_{0\mu_b}$ as follows
\begin{equation} \label{Path:Model:Q0mub}
\begin{aligned}
Q'_{0\mu_b} &= e^{-\mu_a} Q_{0\mu_b}\\
&= 4(1-p_d)^2e^{-\eta_b\mu_b/2-\mu_a}[1-(1-p_d)e^{-\frac14\eta_b\mu_b}]^2. \\
\end{aligned}
\end{equation}
The term $Q'_{0\mu_b}$ appears as an additive term in the key rate formula of Eq.~\eqref{Path:Sim:KeyrateMI} because we assume a forward classical communication (Alice to Bob) is used for postprocessing \cite{Lo:Vacuum:2005,Koashi:Comp:09}. The intuition behind it is that, when Alice sends out a vacuum state as an information carrier, no one (including Eve) can get any information about the final key (Alice's bit) by measuring the signals in the channel.

The overall QBER $E_{\mu_a\mu_b}$ is defined as the error rate in the sifted data. Similar to the derivation of Eq.~\eqref{Path:Model:e11Y11}, due to symmetry, we only need to consider the case of $\theta_a-\theta_b=0$. Without loss of generality, we assume $\theta_a=\theta_b=0$, which leads to $D_{r_0}=D_{s_0}$ and $D_{r_1}=D_{s_1}$ in Eq.~\eqref{Path:Model:Prob4mode}.
In this case, an error happens when the relay announces anticorrelated bits corresponding to clicks on $r_1$-$s_0$ and $r_0$-$s_1$ detectors. The intrinsic error rate, due to background noise and multiphoton states, is then given by
\begin{equation} \label{Path:Model:EQmumuIntr}
\begin{aligned}
E'_{\mu_a\mu_b}Q_{\mu_a\mu_b} &= 2 D_{r_0}(1-D_{r_1})(1-D_{s_0})D_{s_1} \\
&= 2y^2(y-e^{x\cos\Delta_{\phi}})(y-e^{-x\cos\Delta_{\phi}}). \\
\end{aligned}
\end{equation}
It can be verified that Eq.~\eqref{Path:Model:EQmumuIntr} is a decreasing function of $x$. The minimum of $E'_{\mu_a\mu_b}Q_{\mu_a\mu_b}$ is then obtained when $\eta_a\mu_a=\eta_b\mu_b$. Averaging over $\Delta_{\phi}$ in Eq.~\eqref{Path:Model:QmumuGen}, we have
\begin{equation} \label{Path:Model:EQmumuIntrInt}
\begin{aligned}
E'_{\mu_a\mu_b}Q_{\mu_a\mu_b} &= 2y^2[1+y^2-2yI_0(x)], \\
\end{aligned}
\end{equation}
where $x$ and $y$ are defined in Eq.~\eqref{Path:Model:CoherentNotations}. Finally, considering relative-phase distortion errors, in a similar way to Eq.~\eqref{Path:Model:e11Y11}, we obtain
\begin{equation} \label{Path:Model:EQmumu}
\begin{aligned}
E_{\mu_a\mu_b}Q_{\mu_a\mu_b} &= e_0Q_{\mu_a\mu_b}-2(e_0-e_d)y^2[I_0(2x)-1]. \\
\end{aligned}
\end{equation}

\subsection{Phase randomization with postselection} \label{App:Path:PhasePost}
If Alice and Bob randomly set the overall phases of their coherent sources, a large intrinsic QBER is accrued (see Fig.~\ref{Fig:Path:QBER}). From Eq.~\eqref{Path:Model:EQmumuIntr}, the intrinsic QBER is 0 if $\Delta_\phi = 0$, $\mu_a \eta_a = \mu_b \eta_b$, and $p_d =0$. The condition $\Delta_\phi = 0$ implies that Alice and Bob must use the same overall phase value, which jeopardizes the security assumption that requires random-phase values. In order to reduce the cost of error correction, they can, however, inform each other, at the sifting stage, the phase region they used in Eq.~\eqref{Path:Model:PhasePartition}. We remark that this improved data postprocessing is originated from the one proposed in \cite{MXF:DDIQKD:2012}.

Let us take a look at a simple example where Alice sends Bob two-bit classical information for phase postselecting. Then, according to Eq.~\eqref{Path:Model:PhasePartition}, they can divide the phase in $[0,2\pi)$ into four partitions:
\begin{equation} \label{Path:Model:PhasepsPartition2bit}
\begin{aligned}
\left\{[\frac{m\pi}{4},\frac{(m+1)\pi}{4})\cup[\frac{(m+4)\pi}{4},\frac{(m+5)\pi}{4}) \mid m=0,1,2,3\right\}.
\end{aligned}
\end{equation}
The two classical bits for each pulse are used to identify which partition they use for their random phases. Then, the cost of error correction is given by Eq.~\eqref{Path:Sim:KeyrateMI},
\begin{equation} \label{Path:Sim:KeyrateMI2bit}
\begin{aligned}
I_{ec} &= \sum_{m=0}^{3}Q^{m} f H(E^{m}),
\end{aligned}
\end{equation}
where, due to the symmetry, we can assume Alice picks up $m$ from $\{0,1,2,3\}$ randomly and Bob always uses $m=0$.

The gain $Q_{\mu_a\mu_b}$ in Eq.~\eqref{Path:Model:QmumuGen} should be averaged over $\Delta_\phi$ from $m\pi/N$ to $(m+1)\pi/N$, yielding
\begin{equation} \label{Path:Model:QmuIntrps}
\begin{aligned}
Q^{m} &= \frac{N}{\pi}\int_{0}^{\pi/N} d\phi_b \frac{1}{\pi}\int_{m\pi/N}^{(m+1)\pi/N} d\phi_a y^2(e^{-x\cos\Delta_{\phi}}+e^{x\cos\Delta_{\phi}}-2y)^2. \\
\end{aligned}
\end{equation}
Similarly, for the QBER, $E'_{\mu_a\mu_b}Q_{\mu_a\mu_b}$, one should take the integral of Eq.~\eqref{Path:Model:EQmumuIntr} to obtain
\begin{equation} \label{Path:Model:EQmuIntrps}
\begin{aligned}
E'^{m}Q^{m} &= \frac{N}{\pi}\int_{0}^{\pi/N} d\phi_b \frac{1}{\pi}\int_{m\pi/N}^{(m+1)\pi/N} d\phi_a 2y^2(y-e^{x\cos\Delta_{\phi}})(y-e^{-x\cos\Delta_{\phi}}). \\
\end{aligned}
\end{equation}
The intrinsic QBERs for $m=0$ with two cases, $k=2$ and $k=3$, are shown in Fig.~\ref{Fig:Path:QBER}, where $k=\log_2N$.

Let us consider the case when $\eta_a\mu_a=\eta_b\mu_b$, which minimizes the intrinsic QBER of Eq.~\eqref{Path:Model:EQmumuIntr}, hence $\mu'=2\eta_a\mu_a$, $x=\eta_a\mu_a=\mu'/2$, and $y=(1-p_d)e^{-x}$. Assuming $p_d<\mu'\ll1$, and using the first-order approximation to Eq.~\eqref{Path:Model:QmumuGen}, we obtain
\begin{equation} \label{Path:Model:QmuIntrApprox}
\begin{aligned}
Q^{m} 
&= \frac4{N}y^2(1-y)^2+O(\mu'^3), \\
\end{aligned}
\end{equation}
which is independent of $m$, and, for Eq.~\eqref{Path:Model:EQmumuIntr},
\begin{equation} \label{Path:Model:EQmuIntrApprox}
\begin{aligned}
E'^{m}Q^{m} 
&= \frac2{N}y^2(1-y)^2-\frac{2x^2y^3N}{\pi^2} \int_{0}^{\pi/N} d\phi_b \int_{m\pi/N}^{(m+1)\pi/N} d\phi_a  \cos^2\Delta_{\phi} +O(\mu'^3) \\
&= \frac2{N}y^2(1-y)^2 -\frac{x^2y^3}{N} -\frac{x^2y^3N}{4\pi^2} A_{m,N} +O(\mu'^3), \\
\end{aligned}
\end{equation}
where
\begin{equation}
A_{m,N} \equiv -\cos\left[\frac{2 (-1+m) \pi }{N}\right]+2 \cos\left[\frac{2 m \pi }{N}\right]-\cos\left[\frac{2 (1+m) \pi }{N}\right]
\end{equation}
and we use the fact that $1-y=O(\mu')$. From the numerical evaluation, we notice that Eq.~\eqref{Path:Model:EQmuIntrApprox} gives a slightly higher value for QBER than the integral in Eq.~\eqref{Path:Model:EQmuIntrps}. Finally, similar to Eq.~\eqref{Path:Model:EQmumu}, the overall QBER is given by
\begin{equation} \label{Path:Model:EQmuApprox}
\begin{aligned}
E^{m}Q^{m} 
&\approx e_0Q_{\mu_a\mu_b}-(e_0-e_d) \left(\frac{2x^2y^3}{N}+\frac{x^2y^3N}{2\pi^2}A_{m,N}\right). \\
\end{aligned}
\end{equation}


\subsection{Randomized but equal overall phase} \label{Sub:App:Delta0}
Assume Alice and Bob can somehow manage to meet $\Delta_{\phi}=0$. Then, using Eq.~\eqref{Path:Model:SuccProb}, the gain is given by
\begin{equation} \label{Path:Model:GainDelta0}
\begin{aligned}
Q_{\mu_a\mu_b} &= y^2(e^{-x}+e^{x}-2y)^2, \\
\end{aligned}
\end{equation}
and, from Eq.~\eqref{Path:Model:EQmumuIntr}, the corresponding QBER is given by
\begin{equation} \label{Path:Model:QBERDelta0}
\begin{aligned}
E_{\mu_a\mu_b}Q_{\mu_a\mu_b} 
&= e_0 Q_{\mu_a\mu_b}-(e_0-e_d)y^2(e^{x}-e^{-x})^2. \\
\end{aligned}
\end{equation}
One can evaluate the key rate using Eq.~\eqref{Path:Sim:KeyrateMI} by taking the lower bound of $Q'_{0\mu_b}=0$. We numerically verified that the key rate obtained from Eqs.~\eqref{Path:Model:GainDelta0} and \eqref{Path:Model:QBERDelta0} is close to that of the original MDI-QKD scheme in \cite{Lo:MIQKD:2012} for the parameter set given in Table \ref{Tab:Path:SimPara}.

%
%

\subsection{The original MDI-QKD scheme}
In our path-phase encoding scheme of Fig.~\ref{Fig:Path:MIdiag}, the four BB84 states are encoded by the relative phases of two orthogonal optical modes, $r$ and $s$. If we think of single-photon states in $r$ and $s$ modes as a standard basis for qubit representation, our encoding uses the basis vectors of $X$ and $Y$ Pauli operators.
In our setup, one can also encode key information directly onto modes $r$ and $s$ as the third basis ($Z$ basis) for QKD. In fact, the rectilinear basis in the original MDI-QKD can be regarded as using this third basis. The diagonal basis in \cite{Lo:MIQKD:2012} is then equivalent to the $X$ basis in our scheme.

Using the above correspondence, we reproduce the key rate formula for the original MDI-QKD scheme in \cite{Lo:MIQKD:2012}, which is given by
\begin{equation}
\label{Rate:Decoy:Orig}
R \geq Q_{11} [1-H(e_{11})] - Q_\text{rect} f(E_\text{rect}) H(E_\text{rect}),
\end{equation}
where $Q_{11}$ and $e_{11}$ are, respectively, given by Eqs.~\eqref{Path:Model:SourceP11} and ~\eqref{Path:Model:e11Y11}, and $Q_\text{rect}$ and $E_\text{rect}$ are, respectively, the gain and the QBER in the rectilinear basis. The latter two are the only terms that we need to calculate here, as described below. Note that $Q_{11}$ is the same for both rectilinear and diagonal bases.

In the rectilinear basis, Alice chooses one of the two $r$ and $s$ modes and sends a phase-randomized coherent state $\ket{e^{i\phi_a}\sqrt{\mu_a}}$. Similarly, Bob sends $\ket{e^{i\phi_b}\sqrt{\mu_b}}$ in one of the two modes. If different modes are chosen by Alice and Bob, then a click on one of the $r$ detectors, in Fig.~\ref{Fig:Path:MIdiag}, as well as a click on one of the $s$ detectors, correctly indicate the exchange of anticorrelated bits by Alice and Bob. If, however, they choose similar modes, and such a two-click event occurs, they mistakenly assign different bits to their raw keys, and that will be a source of error. The overall gain in the rectilinear basis is then given by the sum of detection probabilities in the above scenarios as follows:
\begin{equation}
\label{Path:Decoy:OriGain}
Q_{\text{rect}} = Q_\text{rect}^{(C)} + Q_\text{rect}^{(E)},
\end{equation}
where
\begin{equation}
Q_\text{rect}^{(C)} = 2(1-p_d)^2e^{-{\mu'}/{2}} \left[1-(1-p_d)e^{-\eta_a\mu_a/2}\right]\left[1-(1-p_d)e^{-\eta_b\mu_b/2}\right]
\end{equation}
represents the detection probability in the first scenario, and
\begin{equation}
Q_\text{rect}^{(E)} = 2 p_d (1-p_d)^2 e^{-{\mu'}/{2}} [I_0(2x)-(1-p_d) e^{-{\mu'}/{2}}]
\end{equation}
represents the detection probability in the second scenario, where $\mu'$ and $x$ are defined in Eq.~\eqref{Path:Model:CoherentNotations}. Note that the above equation also includes averaging over the randomized overall phase.

Finally, considering misalignment errors, we obtain
\begin{equation}
\label{Path:Decoy:OriQBER}
E_\text{rect} Q_\text{rect} = e_d Q_\text{rect}^{(C)} + (1-e_d) Q_\text{rect}^{(E)}.
\end{equation}

\end{appendix}

\bibliographystyle{apsrev4-1}

\bibliography{Bibli}


\end{document}